\documentclass[11pt]{article}
\usepackage{times}
\usepackage{geometry}
\usepackage[utf8]{inputenc}
\usepackage{enumitem,amssymb}
\usepackage{ragged2e}
\newlist{thematic}{itemize}{8}
\setlist[thematic]{label=$\square$}
\usepackage{pifont}
\usepackage{fancyhdr}
\usepackage[USenglish]{babel}
\usepackage[utf8]{inputenc}
\usepackage[T1]{fontenc}
\usepackage{epsfig}
\usepackage{wrapfig}
\usepackage{color}
\usepackage{picinpar}

\usepackage[vflt]{floatflt}
\usepackage{longtable}
\usepackage{caption}
\usepackage{jheppub}   
\usepackage[dvipsnames,svgnames,x11names]{xcolor}
\usepackage[all]{hypcap} 
\usepackage{amssymb,amsmath}
\usepackage{longtable}
\usepackage{amssymb}
\usepackage{amsmath, xspace}
\usepackage{graphics,graphicx} 
\usepackage{rotating}
\usepackage{color}
\usepackage{xcolor}
\usepackage{multirow}
\usepackage{subfigure}
\usepackage{sectsty}
\usepackage[outercaption]{sidecap}
\sidecaptionvpos{figure}{c}

\usepackage{multicol}

\usepackage{soul}

\usepackage[
  backend=biber,
  style=numeric,
  sorting=none,
  maxcitenames=1,
  maxbibnames=2,
  giveninits=true  
]{biblatex}

\DeclareNameAlias{default}{family-given}

\AtEveryBibitem{%
  \clearfield{title}%
  \clearfield{doi}%
  \clearfield{url}%
  \clearfield{eprint}%
  \clearfield{isbn}%
  \clearfield{issn}%
  \clearlist{language}%
}

\renewbibmacro{in:}{}

\RequireBibliographyStyle{numeric} 
\DeclareFieldFormat{pages}{%
  \ifdefstring{\thefield{pages}}{}{}{%
    \mkfirstpage{\thefield{pages}}%
  }%
}

\DeclareFieldFormat{year}{#1}

\DeclareFieldFormat[article]{volume}{#1}

\DeclareFieldFormat[article]{number}{}

\DeclareBibliographyDriver{article}{%
  \printnames{author}%
  \setunit{\addcomma\space}%
  \printfield{year}%
  \setunit{\addcomma\space}%
  \printfield{journaltitle}%
  \addspace%
  \printfield{volume}%
  \setunit{\addcomma\space}%
  \printfield{pages}%
  \finentry
}

\usepackage{enumitem}
\setlist[enumerate]{itemsep=0pt, parsep=0pt}
\setlist[itemize]{itemsep=0pt, parsep=0pt}

\definecolor{DarkGreen}{rgb}{0.0, 0.3, 0.0}
\definecolor{purple}{rgb}{0.5, 0.0, 0.5}
\definecolor{red}{rgb}{1, 0.0, 0.0}
\definecolor{green}{rgb}{0, 1.0, 0.0}

\def\3he{$^3{\rm He}$}

\def\lsim{\mathrel{\lower2.5pt\vbox{\lineskip=0pt\baselineskip=0pt
           \hbox{$<$}\hbox{$\sim$}}}}

\def\gsim{\mathrel{\lower2.5pt\vbox{\lineskip=0pt\baselineskip=0pt
           \hbox{$>$}\hbox{$\sim$}}}}

\begin{document}
\raggedright
\huge
Serendipitous and targeted mm/sub-mm transient searches with wide-FOV telescope  
\linebreak
\normalsize

%
%
\vspace{-0.6cm}
\thispagestyle{empty}
\noindent
\vspace{1em}
\begin{flushleft}
\textbf{Authors:}
\vspace{1em}

\begin{tabular}{p{0.2cm} l p{1cm} l}
& \textbf{Karri Koljonen}$^1$    &  & \textit{Finnish Centre for Astronomy with ESO (FINCA),} \\
& & & \textit{University of Turku, Finland \& NTNU, Norway} \\
& \textbf{Claudio Ricci}    &  &  \textit{University of Geneva, Switzerland} \\
& \textbf{Thomas Stanke}  &  & \textit{Max-Planck-Institute for Extraterrestrial Physics, Germany} \\ \\
& \multicolumn{3}{p{14cm}}{\it
\textbf{Doug Johnstone} (NRC Herzberg Astronomy and Astrophysics, Canada)
\textbf{Atul Mohan} (NASA Goddard Space Flight Center \& Catholic University of America, US)
\textbf{Francisco M. Montenegro-Montes} (Institute of Particle and Cosmos Physics, Univ. Complutense de Madrid, Spain)
\textbf{John Orlowski-Scherer} (University of Pennsylvania, US)

}
\end{tabular}
\end{flushleft}
\footnotetext[1]{\url{karri.koljonen@utu.fi}}

\vfill

 \captionsetup{labelformat=empty}
\begin{figure}[h]
   \centering
\includegraphics[width=.65\textwidth]{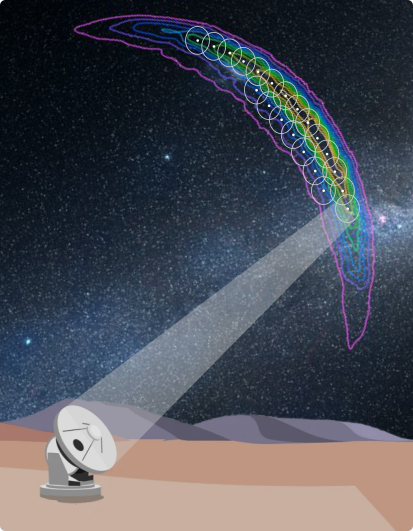}
   \caption{}
\end{figure}
\vspace{-15mm}

\setcounter{figure}{0}
\captionsetup{labelformat=default}

\vfill
\vspace{0.6cm}
\textbf{Science Keywords}: \textit{methods: observational — telescopes — submillimetre: general — radio continuum: transients — stars: protostars — stars: flare — galaxies: jets — gravitational waves — neutrinos — instrumentation: polarimeters — surveys} 
\linebreak

\justify

\pagebreak


\section*{Abstract}

The millimeter/sub-millimeter (mm/sub-mm) sky remains a rich but under-explored frontier for transient and variable phenomena. A wide-field, high-sensitivity instrument with a large aperture and degree-scale field of view would open this regime, enabling both systematic survey monitoring and rapid-response follow-up. Key science opportunities include Galactic Plane monitoring and surveys to discover and characterize time-variable emission from young stellar objects, magnetically active and flaring stars, compact binaries, and explosive events, as well as prompt responses to multi-messenger alerts with large localization regions (e.g., gravitational-wave triggers). Multi-band capability, rapid slewing, and high sensitivity are essential to probe energetic processes such as jet launching, relativistic shocks and accretion flows in unprecedented detail. While long‑term monitoring is well established at radio and optical/infrared wavelengths, mm/sub-mm observations uniquely bridge the spectral gap between these regimes, directly probing obscured environments that are inaccessible elsewhere. Large-scale monitoring programs will yield legacy datasets crucial for population studies through the 2040s and beyond.

\section{Scientific context and motivation}


Astrophysical transients trace the Universe’s most energetic processes from the deaths of stars and birth of compact objects to jet launching, particle acceleration, and multi-messenger events probing extreme physical conditions beyond any laboratory experiment. Yet the mm/sub-mm time-domain window remains largely under-explored despite key advantages over other wavelengths \cite{Eftekhari2022}.

\vspace{0.1cm} \noindent The 2030s–2040s will be a golden age for time-domain astronomy driven by the wide-field surveys across the electromagnetic spectrum (such as Vera C. Rubin Observatory, Roman Space Telescope, Ultraviolet Transient Astronomy Satellite, Cherenkov Telescope Array Observatory, Square Kilometre Array Observatory, and Transient High-Energy Sky and Early Universe Surveyor) and advanced multi-messenger detectors in ice (e.g., IceCube-Gen2), sea (e.g., The Cubic Kilometre Neutrino Telescope, and The tRopIcal DEep-sea Neutrino Telescope) and space (e.g., The Laser Interferometer Space Antenna). A glaring omission from this landscape is a high-sensitivity next-generation mm/sub-mm observatory designed for time-domain studies. While upcoming observatories such as The Fred Young Submillimeter Telescope will offer valuable pathfinding capabilities, their collecting area and sensitivity remain insufficient to fully address the most demanding transient science cases. Yet, mm/sub-mm wavelengths are uniquely powerful for revealing embedded or dust-obscured transients probing emission that escapes dense environments in star-forming regions \cite{Francis2022}, dust-enshrouded galactic nuclei \cite{delPalacio2025}, and early-phase supernova ejecta \cite{Maeda2021}. Early relativistic shocks in the jets of gamma-ray bursts (GRBs), fast blue optical transients (FBOTs), tidal disruption events (TDEs), and rapid accretion episodes onto supermassive black holes (e.g., changing‑look AGN) often peak at mm frequencies, providing diagnostics of magnetic fields, particle acceleration, and jet energetics \cite{Laskar2018,Laskar2019,Ho2019,Alexander2020}. In addition, shocks in the ejecta of novae \cite{Chomiuk2021}, core-collapse supernovae \cite{Berger2023}, and Galactic transients \cite{Tetarenko2019} produce strong and rapidly evolving mm/sub-mm signatures. Compact object mergers that emit gravitational waves (GWs) and neutrinos can generate isotropic or mildly beamed mm emission, enabling robust electromagnetic association even for off-axis or heavily obscured events \cite{Alexander2017,Eftekhari2022}.

\vspace{0.1cm} \noindent Despite the rich physics accessible at mm/sub-mm wavelengths, the field remains discovery-limited. Wide field-of-view (FOV) cosmic microwave background (CMB) telescopes can detect bright mm sources serendipitously, but their sensitivity, cadence, and angular resolution are not optimized for astrophysical transients \cite{Guns2021,HerviasCaimapo2024}. Interferometers such as ALMA or NOEMA provide exquisite sensitivity for targeted follow-up but lack the FOV and survey speed required for blind discovery. At present, no facility combines arcsecond-scale resolution, sub-mJy sensitivity, and a degree-scale FOV. Addressing this gap, and realizing the rich scientific potential it presents, motivates the development of a next-generation, large single-dish mm/sub-mm telescope such as AtLAST \cite{Mroczkowski2025}.

\pagebreak


\begin{figure*}[h]
\centering
\begin{minipage}{0.48\textwidth}
\section{Science case}
A next‑generation wide-FOV mm/sub‑mm telescope would transform our understanding of transient and variable astrophysical phenomena across Galactic and extragalactic environments. Its high sensitivity and rapid‑response capabilities would place it at the center of multi‑messenger astrophysics in the 2040s \cite{Miller2024}. Sources detected via GW or neutrino observatories typically have coarse sky localizations ($\sim10-100$ deg$^2$) necessitating rapid and efficient tiling of these regions to identify electromagnetic counterparts (see Fig. 1). Rapid mm/sub‑mm follow‑up is essential because emission from relativistic shocks and jet–cocoon interactions often peaks hours to days after a trigger, before fading on timescales of days to weeks---more rapidly than radio but often outlasting optical or X‑ray signals. Observations in this regime probe early‑time conditions such as reverse shocks \cite{Laskar2019,Eftekhari2022,deWet2024} and optically thin radiation at jet bases \cite{Hovatta2019} offering diagnostics unique and complementary to other wavelengths. Rapid-response capabilities and automatic triggering pipelines (e.g., via VOEvent or similar real‑time protocols) are essential in locating and characterizing multi-messenger sources before critical early‑time emission fades.
 
\end{minipage}
\hfill
\begin{minipage}{0.48\textwidth}
\includegraphics[width=\linewidth]{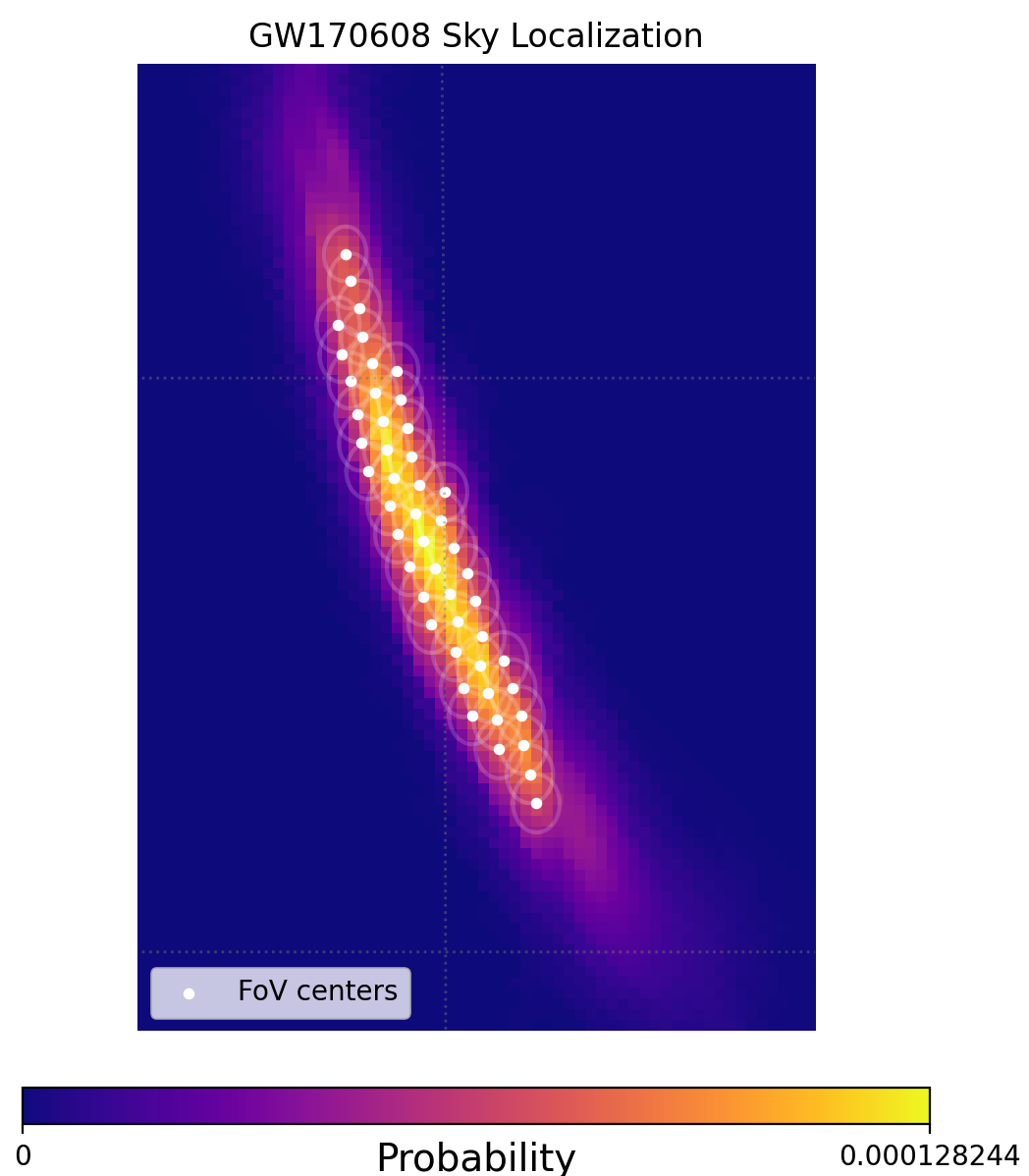}
\caption{\small \it Example of mapping GW sky localization regions. The figure shows LIGO/Virgo event GW170608 from GWTC-1, with circular 2-degree-diameter telescope footprints optimized for highest-probability coverage. 44 fields covering 64 deg$^2$ capture 40\% of the probability region, mapped with $\sim$0.1 mJy/beam sensitivity in 3 hours (Band 3; 91.5 GHz central frequency, 51 GHz bandwidth, 10 s per field). Observations progress from the highest- to lowest-probability areas with $\sim$2 deg/s slews.}
\end{minipage}
\end{figure*}

\vspace{-0.3cm} \noindent The same capabilities would unlock a rich landscape of Galactic time‑variable phenomena. The Galactic plane is populated by deeply embedded young stellar objects, compact binaries, and magnetically active stars, many of which remain invisible or heavily extincted in the optical and infrared. In protostellar cores, episodic accretion bursts causes the dust continuum to brighten rapidly producing a time-variable signature \cite{Lee2021,Mairs2024}. Repeated monitoring of wide fields at arcsecond resolution would provide direct statistics on the frequency, duration, and energetics of such accretion outbursts. Magnetically active stars produce non-thermal flares detectable at mm wavelengths, offering insight into particle acceleration and magnetic reconnection in stellar magnetospheres \cite{Mairs2019,MacGregor2020,MacGregor2021,Lovell2024}. A wide‑field Stokes-V polarimetric, simultaneous multi‑band survey would uniquely diagnose the particle‑acceleration and plasma‑heating processes driving these events, revealing the underlying magnetic geometry and emission mechanisms \cite{Callingham2023}. Coordinated multi‑wavelength observations would establish the most comprehensive view yet of stellar magnetic activity, providing a prototype for next‑generation polarimetric time‑domain surveys in the mm/sub‑mm regime. Multi‑epoch surveys extended over years---with regular monthly or weekly cadence---would be needed to measure long‑term accretion and magnetic‑cycle variability establishing baseline light curves for a broad range of young and evolved stellar populations.

\vspace{0.1cm} \noindent A wide-FOV also enables the discovery of extragalactic transients. GRBs, FBOTs and TDEs produce synchrotron emission that often peaks at mm/sub-mm wavelengths. GRB reverse shocks can produce bright mm flashes within hours to days of the explosion (Fig. 2). Orphan afterglows, which lack a detected high-energy trigger, may only be detectable in the sub-mm \cite{Eftekhari2022}. Some FBOTs exhibit extraordinarily bright sub-mm emission  \cite{Ho2019,Perley2019}. TDEs can produce mm emission from a jet or from reprocessed radiation in the circumnuclear medium \cite{Alexander2020}. Operating continuous, wide-area surveys would enable detection and tracking of these sources, construction of light curves, and characterization of their spectral evolution through multi-band observations.

\vspace{0.1cm} \noindent Beyond discovery and rapid follow-up, long-term observing campaigns and a wide-FOV will produce a legacy archive of the variable sub-mm sky. Observations acquired for other science programs can be used as serendipitous survey data to search for variability. This will yield a rich database of mm/sub-mm light curves, flux-variability statistics, and transient populations that will remain invaluable for decades.

\begin{figure*}[h]
\centering
\begin{minipage}{0.48\textwidth}
\section{Technical requirements}
To meet the above science goals, several key technical features are required. Scheduling must be highly flexible: the facility should seamlessly combine routine survey operations with rapid‑response overrides triggered by external alerts. The system must support bidirectional real‑time communication—both receiving and issuing automated triggers. Rapid slewing capability ($\sim$deg/s) enables on‑source observations within minutes of the trigger. In addition, a dedicated real‑time transient pipeline is essential. This system must handle calibrated imaging, difference imaging, and automated candidate detection. Machine-learning-based classifiers will be required to distinguish astrophysical transients from atmospheric fluctuations, instrumental artifacts, and moving foreground sources \cite{Moller2021,Sanchez2019}.
\end{minipage}
\hfill
\begin{minipage}{0.48\textwidth}
\includegraphics[width=\linewidth]{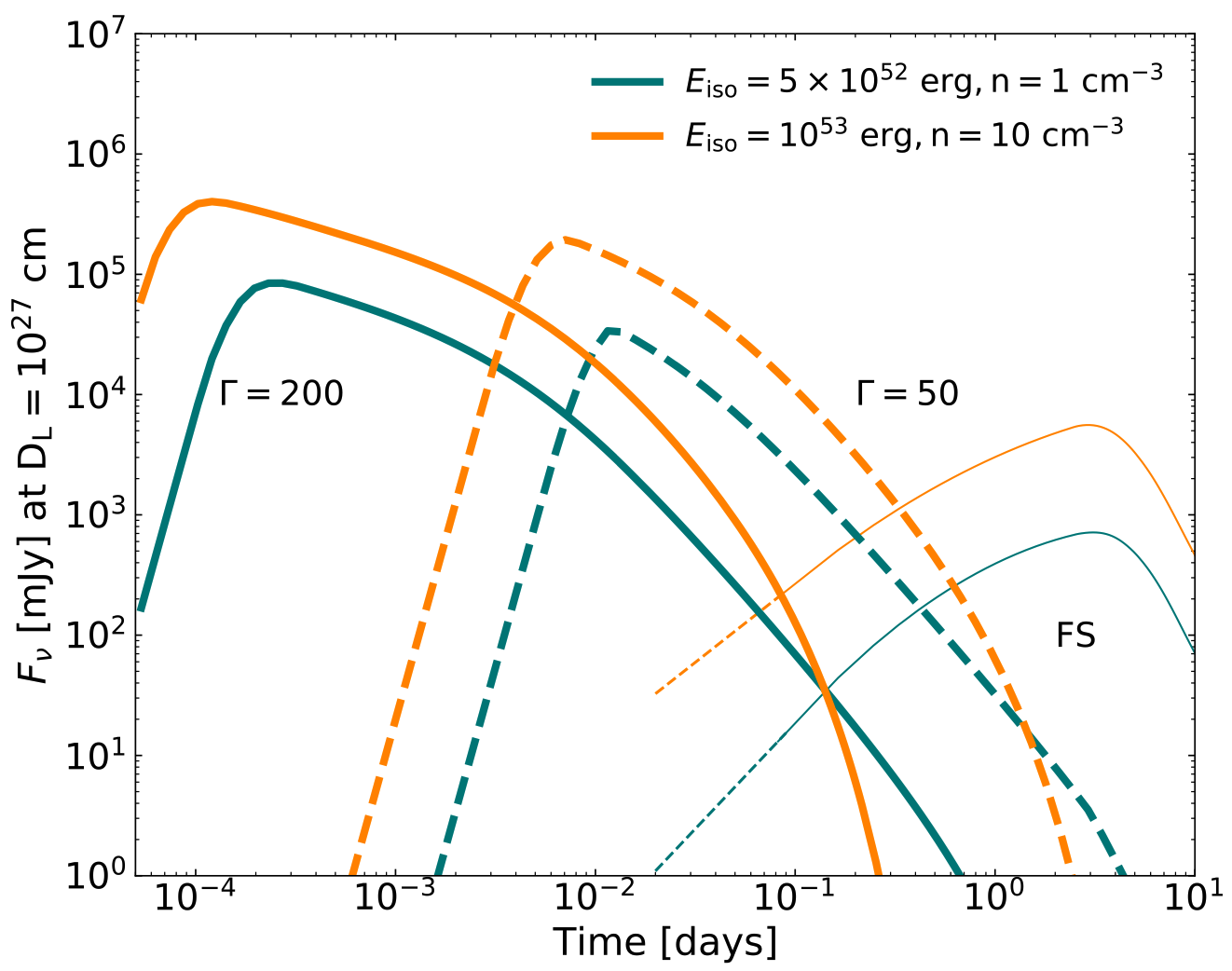}
\caption{\small \it Reverse shock models at 100 GHz for the GRB jets \cite{Eftekhari2022}. Solid and dashed lines correspond to initial Lorentz factors of $\Gamma$ = 200 and 50, respectively. Thin lines correspond to the forward shock component which peaks at later times.}
\end{minipage}
\end{figure*}

\vspace{-0.3cm} \noindent The instrument must deliver a large instantaneous FOV ($\geq$1 deg$^2$) enabling efficient survey mapping and rapid tiling of large error regions. Arcsecond angular resolution is necessary to allow precise localization of sources and enable efficient multi-wavelength follow-up. Broad frequency coverage from 30--950 GHz is highly desirable, with wide instantaneous bandwidth in each band to maximize continuum sensitivity and enable rapid spectral characterization of detected transients. Sub-minute integrations and variable cadences (hours–weeks) are needed to track fast-evolving transients. On-the-fly scanning is optimal for providing uniform coverage with minimal overhead. Full‑Stokes polarimetric capability is essential not only to distinguish synchrotron, gyrosynchrotron, and coherent stellar emission from thermal dust, but also to capture polarization and spectral‑index evolution indicative of magnetic‑reconnection and particle‑acceleration processes. Simultaneous multi‑frequency polarimetry will be the key enabler for stellar‑flare searches and for characterizing magnetic‑field geometries in diverse transient sources.

\begin{multicols}{2}
\printbibliography

@ARTICLE{Mroczkowski2025,
       author = {{Mroczkowski}, Tony and {Gallardo}, Patricio A. and {Timpe}, Martin and {Kiselev}, Aleksej and {Groh}, Manuel and {Kaercher}, Hans and {Reichert}, Matthias and {Cicone}, Claudia and {Puddu}, Roberto and {Dubois-dit-Bonclaude}, Pierre and {Bok}, Daniel and {Dahl}, Erik and {Macintosh}, Mike and {Dicker}, Simon and {Viole}, Isabelle and {Sartori}, Sabrina and {Valenzuela Venegas}, Guillermo Andr{\'e}s and {Zeyringer}, Marianne and {Niemack}, Michael and {Poppi}, Sergio and {Olguin}, Rodrigo and {Hatziminaoglou}, Evanthia and {De Breuck}, Carlos and {Klaassen}, Pamela and {Montenegro-Montes}, Francisco Miguel and {Zimmerer}, Thomas},
        title = "{The conceptual design of the 50-meter Atacama Large Aperture Submillimeter Telescope (AtLAST)}",
      journal = {\aap},
         year = 2025,
        month = feb,
       volume = {694},
          eid = {A142},
        pages = {A142},
          doi = {10.1051/0004-6361/202449786},
archivePrefix = {arXiv},
       eprint = {2402.18645},
 primaryClass = {astro-ph.IM},
       adsurl = {https://ui.adsabs.harvard.edu/abs/2025A&A...694A.142M}
}

@ARTICLE{Guns2021,
       author = {{Guns}, S. and {Foster}, A. and {Daley}, C. and {Rahlin}, A. and {Whitehorn}, N. and {Ade}, P.~A.~R. and {Ahmed}, Z. and {Anderes}, E. and {Anderson}, A.~J. and {Archipley}, M. and {Avva}, J.~S. and {Aylor}, K. and {Balkenhol}, L. and {Barry}, P.~S. and {Basu Thakur}, R. and {Benabed}, K. and {Bender}, A.~N. and {Benson}, B.~A. and {Bianchini}, F. and {Bleem}, L.~E. and {Bouchet}, F.~R. and {Bryant}, L. and {Byrum}, K. and {Carlstrom}, J.~E. and {Carter}, F.~W. and {Cecil}, T.~W. and {Chang}, C.~L. and {Chaubal}, P. and {Chen}, G. and {Cho}, H.-M. and {Chou}, T.-L. and {Cliche}, J.-F. and {Crawford}, T.~M. and {Cukierman}, A. and {de Haan}, T. and {Denison}, E.~V. and {Dibert}, K. and {Ding}, J. and {Dobbs}, M.~A. and {Dutcher}, D. and {Everett}, W. and {Feng}, C. and {Ferguson}, K.~R. and {Fu}, J. and {Galli}, S. and {Gambrel}, A.~E. and {Gardner}, R.~W. and {Goeckner-Wald}, N. and {Gualtieri}, R. and {Gupta}, N. and {Guyser}, R. and {Halverson}, N.~W. and {Harke-Hosemann}, A.~H. and {Harrington}, N.~L. and {Henning}, J.~W. and {Hilton}, G.~C. and {Hivon}, E. and {Holder}, G.~P. and {Holzapfel}, W.~L. and {Hood}, J.~C. and {Howe}, D. and {Huang}, N. and {Irwin}, K.~D. and {Jeong}, O.~B. and {Jonas}, M. and {Jones}, A. and {Khaire}, T.~S. and {Knox}, L. and {Kofman}, A.~M. and {Korman}, M. and {Kubik}, D.~L. and {Kuhlmann}, S. and {Kuo}, C.-L. and {Lee}, A.~T. and {Leitch}, E.~M. and {Lowitz}, A.~E. and {Lu}, C. and {Marrone}, D.~P. and {Meyer}, S.~S. and {Michalik}, D. and {Millea}, M. and {Montgomery}, J. and {Nadolski}, A. and {Natoli}, T. and {Nguyen}, H. and {Noble}, G.~I. and {Novosad}, V. and {Omori}, Y. and {Padin}, S. and {Pan}, Z. and {Paschos}, P. and {Pearson}, J. and {Phadke}, K.~A. and {Posada}, C.~M. and {Prabhu}, K. and {Quan}, W. and {Reichardt}, C.~L. and {Riebel}, D. and {Riedel}, B. and {Rouble}, M. and {Ruhl}, J.~E. and {Sayre}, J.~T. and {Schiappucci}, E. and {Shirokoff}, E. and {Smecher}, G. and {Sobrin}, J.~A. and {Stark}, A.~A. and {Stephen}, J. and {Story}, K.~T. and {Suzuki}, A. and {Thompson}, K.~L. and {Thorne}, B. and {Tucker}, C. and {Umilta}, C. and {Vale}, L.~R. and {Vieira}, J.~D. and {Wang}, G. and {Wu}, W.~L.~K. and {Yefremenko}, V. and {Yoon}, K.~W. and {Young}, M.~R. and {Zhang}, L.},
        title = "{Detection of Galactic and Extragalactic Millimeter-wavelength Transient Sources with SPT-3G}",
      journal = {\apj},
         year = 2021,
        month = aug,
       volume = {916},
       number = {2},
          eid = {98},
        pages = {98},
          doi = {10.3847/1538-4357/ac06a3},
archivePrefix = {arXiv},
       eprint = {2103.06166},
 primaryClass = {astro-ph.HE},
       adsurl = {https://ui.adsabs.harvard.edu/abs/2021ApJ...916...98G}
}

@ARTICLE{Eftekhari2022,
       author = {{Eftekhari}, T. and {Berger}, E. and {Metzger}, B.~D. and {Laskar}, T. and {Villar}, V.~A. and {Alexander}, K.~D. and {Holder}, G.~P. and {Vieira}, J.~D. and {Whitehorn}, N. and {Williams}, P.~K.~G.},
        title = "{Extragalactic Millimeter Transients in the Era of Next-generation CMB Surveys}",
      journal = {\apj},
         year = 2022,
        month = aug,
       volume = {935},
       number = {1},
          eid = {16},
        pages = {16},
          doi = {10.3847/1538-4357/ac7ce8},
archivePrefix = {arXiv},
       eprint = {2110.05494},
 primaryClass = {astro-ph.HE},
       adsurl = {https://ui.adsabs.harvard.edu/abs/2022ApJ...935...16E}
}

@ARTICLE{Lee2021,
       author = {{Lee}, Yong-Hee and {Johnstone}, Doug and {Lee}, Jeong-Eun and {Herczeg}, Gregory and {Mairs}, Steve and {Contreras-Pe{\~n}a}, Carlos and {Hatchell}, Jennifer and {Naylor}, Tim and {Bell}, Graham S. and {Bourke}, Tyler L. and {Broughton}, Colton and {Francis}, Logan and {Gupta}, Aashish and {Harsono}, Daniel and {Liu}, Sheng-Yuan and {Park}, Geumsook and {Plovie}, Spencer and {Moriarty-Schieven}, Gerald H. and {Scholz}, Aleks and {Sharma}, Tanvi and {Teixeira}, Paula Stella and {Wang}, Yao-Te and {Aikawa}, Yuri and {Bower}, Geoffrey C. and {Vivien Chen}, Huei-Ru and {Bae}, Jaehan and {Baek}, Giseon and {Chapman}, Scott and {Ping Chen}, Wen and {Du}, Fujun and {Dutta}, Somnath and {Forbrich}, Jan and {Guo}, Zhen and {Inutsuka}, Shu-ichiro and {Kang}, Miju and {Kirk}, Helen and {Kuan}, Yi-Jehng and {Kwon}, Woojin and {Lai}, Shih-Ping and {Lalchand}, Bhavana and {Lane}, James M.~M. and {Lee}, Chin-Fei and {Liu}, Tie and {Morata}, Oscar and {Pearson}, Samuel and {Pon}, Andy and {Sahu}, Dipen and {Shang}, Hsien and {Stamatellos}, Dimitris and {Tang}, Shih-Yun and {Xu}, Ziyan and {Yoo}, Hyunju and {Rawlings}, Jonathan M.~C.},
        title = "{The JCMT Transient Survey: Four-year Summary of Monitoring the Submillimeter Variability of Protostars}",
      journal = {\apj},
         year = 2021,
        month = oct,
       volume = {920},
       number = {2},
          eid = {119},
        pages = {119},
          doi = {10.3847/1538-4357/ac1679},
archivePrefix = {arXiv},
       eprint = {2107.10750},
 primaryClass = {astro-ph.SR},
       adsurl = {https://ui.adsabs.harvard.edu/abs/2021ApJ...920..119L}
}

@ARTICLE{Mairs2024,
       author = {{Mairs}, Steve and {Lee}, Seonjae and {Johnstone}, Doug and {Broughton}, Colton and {Lee}, Jeong-Eun and {Herczeg}, Gregory J. and {Bell}, Graham S. and {Chen}, Zhiwei and {Contreras-Pe{\~n}a}, Carlos and {Francis}, Logan and {Hatchell}, Jennifer and {Kim}, Mi-Ryang and {Liu}, Sheng-Yuan and {Park}, Geumsook and {Qiu}, Keping and {Wang}, Yao-Te and {Zhang}, Xu and {JCMT Transient Team}},
        title = "{The JCMT Transient Survey: Six Year Summary of 450/850 {\ensuremath{\mu}}m Protostellar Variability and Calibration Pipeline Version 2.0}",
      journal = {\apj},
         year = 2024,
        month = may,
       volume = {966},
       number = {2},
          eid = {215},
        pages = {215},
          doi = {10.3847/1538-4357/ad35b6},
archivePrefix = {arXiv},
       eprint = {2401.03549},
 primaryClass = {astro-ph.IM},
       adsurl = {https://ui.adsabs.harvard.edu/abs/2024ApJ...966..215M}
}

@ARTICLE{deWet2024,
       author = {{de Wet}, Simon and {Laskar}, Tanmoy and {Groot}, Paul J. and {Barniol Duran}, Rodolfo and {Berger}, Edo and {Bhandari}, Shivani and {Eftekhari}, Tarraneh and {Guidorzi}, Cristiano and {Kobayashi}, Shiho and {Perley}, Daniel A. and {Sari}, Re'em and {Schroeder}, Genevieve},
        title = "{A Millimeter Rebrightening in GRB 210702A}",
      journal = {\apj},
         year = 2024,
        month = oct,
       volume = {974},
       number = {2},
          eid = {279},
        pages = {279},
          doi = {10.3847/1538-4357/ad77bb},
archivePrefix = {arXiv},
       eprint = {2408.14641},
 primaryClass = {astro-ph.HE},
       adsurl = {https://ui.adsabs.harvard.edu/abs/2024ApJ...974..279D}
}

@ARTICLE{HerviasCaimapo2024,
       author = {{Herv{\'\i}as-Caimapo}, Carlos and {Naess}, Sigurd and {Hincks}, Adam D. and {Calabrese}, Erminia and {Devlin}, Mark J. and {Dunkley}, Jo and {D{\"u}nner}, Rolando and {Gallardo}, Patricio A. and {Hilton}, Matt and {Ho}, Anna Y.~Q. and {Huffenberger}, Kevin M. and {Ma}, Xiaoyi and {Madhavacheril}, Mathew S. and {Niemack}, Michael D. and {Orlowski-Scherer}, John and {Page}, Lyman A. and {Partridge}, Bruce and {Puddu}, Roberto and {Salatino}, Maria and {Sif{\'o}n}, Crist{\'o}bal and {Staggs}, Suzanne T. and {Vargas}, Cristian and {Vavagiakis}, Eve M. and {Wollack}, Edward J.},
        title = "{The Atacama cosmology telescope: flux upper limits from a targeted search for extragalactic transients}",
      journal = {\mnras},
         year = 2024,
        month = apr,
       volume = {529},
       number = {3},
        pages = {3020},
          doi = {10.1093/mnras/stae583},
archivePrefix = {arXiv},
       eprint = {2301.07651},
 primaryClass = {astro-ph.HE},
       adsurl = {https://ui.adsabs.harvard.edu/abs/2024MNRAS.529.3020H}
}

@ARTICLE{Ho2019,
       author = {{Ho}, Anna Y.~Q. and {Phinney}, E. Sterl and {Ravi}, Vikram and {Kulkarni}, S.~R. and {Petitpas}, Glen and {Emonts}, Bjorn and {Bhalerao}, V. and {Blundell}, Ray and {Cenko}, S. Bradley and {Dobie}, Dougal and {Howie}, Ryan and {Kamraj}, Nikita and {Kasliwal}, Mansi M. and {Murphy}, Tara and {Perley}, Daniel A. and {Sridharan}, T.~K. and {Yoon}, Ilsang},
        title = "{AT2018cow: A Luminous Millimeter Transient}",
      journal = {\apj},
         year = 2019,
        month = jan,
       volume = {871},
       number = {1},
          eid = {73},
        pages = {73},
          doi = {10.3847/1538-4357/aaf473},
archivePrefix = {arXiv},
       eprint = {1810.10880},
 primaryClass = {astro-ph.HE},
       adsurl = {https://ui.adsabs.harvard.edu/abs/2019ApJ...871...73H}
}

@ARTICLE{Laskar2018,
       author = {{Laskar}, Tanmoy and {Alexander}, Kate D. and {Berger}, Edo and {Guidorzi}, Cristiano and {Margutti}, Raffaella and {Fong}, Wen-fai and {Kilpatrick}, Charles D. and {Milne}, Peter and {Drout}, Maria R. and {Mundell}, C.~G. and {Kobayashi}, Shiho and {Lunnan}, Ragnhild and {Barniol Duran}, Rodolfo and {Menten}, Karl M. and {Ioka}, Kunihito and {Williams}, Peter K.~G.},
        title = "{First ALMA Light Curve Constrains Refreshed Reverse Shocks and Jet Magnetization in GRB 161219B}",
      journal = {\apj},
         year = 2018,
        month = aug,
       volume = {862},
       number = {2},
          eid = {94},
        pages = {94},
          doi = {10.3847/1538-4357/aacbcc},
archivePrefix = {arXiv},
       eprint = {1808.09476},
 primaryClass = {astro-ph.HE},
       adsurl = {https://ui.adsabs.harvard.edu/abs/2018ApJ...862...94L}
}

@ARTICLE{Laskar2019,
       author = {{Laskar}, Tanmoy and {van Eerten}, Hendrik and {Schady}, Patricia and {Mundell}, C.~G. and {Alexander}, Kate D. and {Barniol Duran}, Rodolfo and {Berger}, Edo and {Bolmer}, J. and {Chornock}, Ryan and {Coppejans}, Deanne L. and {Fong}, Wen-fai and {Gomboc}, Andreja and {Jordana-Mitjans}, N{\'u}ria and {Kobayashi}, Shiho and {Margutti}, Raffaella and {Menten}, Karl M. and {Sari}, Re'em and {Yamazaki}, Ryo and {Lipunov}, V.~M. and {Gorbovskoy}, E. and {Kornilov}, V.~G. and {Tyurina}, N. and {Zimnukhov}, D. and {Podesta}, R. and {Levato}, H. and {Buckley}, D.~A.~H. and {Tlatov}, A. and {Rebolo}, R. and {Serra-Ricart}, M.},
        title = "{A Reverse Shock in GRB 181201A}",
      journal = {\apj},
         year = 2019,
        month = oct,
       volume = {884},
       number = {2},
          eid = {121},
        pages = {121},
          doi = {10.3847/1538-4357/ab40ce},
archivePrefix = {arXiv},
       eprint = {1907.13128},
 primaryClass = {astro-ph.HE},
       adsurl = {https://ui.adsabs.harvard.edu/abs/2019ApJ...884..121L}
}

@ARTICLE{Perley2019,
       author = {{Perley}, Daniel A. and {Mazzali}, Paolo A. and {Yan}, Lin and {Cenko}, S. Bradley and {Gezari}, Suvi and {Taggart}, Kirsty and {Blagorodnova}, Nadia and {Fremling}, Christoffer and {Mockler}, Brenna and {Singh}, Avinash and {Tominaga}, Nozomu and {Tanaka}, Masaomi and {Watson}, Alan M. and {Ahumada}, Tom{\'a}s and {Anupama}, G.~C. and {Ashall}, Chris and {Becerra}, Rosa L. and {Bersier}, David and {Bhalerao}, Varun and {Bloom}, Joshua S. and {Butler}, Nathaniel R. and {Copperwheat}, Chris and {Coughlin}, Michael W. and {De}, Kishalay and {Drake}, Andrew J. and {Duev}, Dmitry A. and {Frederick}, Sara and {Gonz{\'a}lez}, J. Jes{\'u}s and {Goobar}, Ariel and {Heida}, Marianne and {Ho}, Anna Y.~Q. and {Horst}, John and {Hung}, Tiara and {Itoh}, Ryosuke and {Jencson}, Jacob E. and {Kasliwal}, Mansi M. and {Kawai}, Nobuyuki and {Khanam}, Tanazza and {Kulkarni}, Shrinivas R. and {Kumar}, Brajesh and {Kumar}, Harsh and {Kutyrev}, Alexander S. and {Lee}, William H. and {Maeda}, Keiichi and {Mahabal}, Ashish and {Murata}, Katsuhiro L. and {Neill}, James D. and {Ngeow}, Chow-Choong and {Penprase}, Bryan and {Pian}, Elena and {Quimby}, Robert and {Ramirez-Ruiz}, Enrico and {Richer}, Michael G. and {Rom{\'a}n-Z{\'u}{\~n}iga}, Carlos G. and {Sahu}, D.~K. and {Srivastav}, Shubham and {Socia}, Quentin and {Sollerman}, Jesper and {Tachibana}, Yutaro and {Taddia}, Francesco and {Tinyanont}, Samaporn and {Troja}, Eleonora and {Ward}, Charlotte and {Wee}, Jerrick and {Yu}, Po-Chieh},
        title = "{The fast, luminous ultraviolet transient AT2018cow: extreme supernova, or disruption of a star by an intermediate-mass black hole?}",
      journal = {\mnras},
         year = 2019,
        month = mar,
       volume = {484},
       number = {1},
        pages = {1031},
          doi = {10.1093/mnras/sty3420},
archivePrefix = {arXiv},
       eprint = {1808.00969},
 primaryClass = {astro-ph.HE},
       adsurl = {https://ui.adsabs.harvard.edu/abs/2019MNRAS.484.1031P}
}

@ARTICLE{Alexander2020,
       author = {{Alexander}, Kate D. and {van Velzen}, Sjoert and {Horesh}, Assaf and {Zauderer}, B. Ashley},
        title = "{Radio Properties of Tidal Disruption Events}",
      journal = {\ssr},
         year = 2020,
        month = jun,
       volume = {216},
       number = {5},
          eid = {81},
        pages = {81},
          doi = {10.1007/s11214-020-00702-w},
archivePrefix = {arXiv},
       eprint = {2006.01159},
 primaryClass = {astro-ph.HE},
       adsurl = {https://ui.adsabs.harvard.edu/abs/2020SSRv..216...81A}
}

@ARTICLE{Sanchez2019,
       author = {{S{\'a}nchez}, B. and {Dom{\'\i}nguez R.}, M.~J. and {Lares}, M. and {Beroiz}, M. and {Cabral}, J.~B. and {Gurovich}, S. and {Qui{\~n}ones}, C. and {Artola}, R. and {Colazo}, C. and {Schneiter}, M. and {Girardini}, C. and {Tornatore}, M. and {Nilo Castell{\'o}n}, J.~L. and {Garc{\'\i}a Lambas}, D. and {D{\'\i}az}, M.~C.},
        title = "{Machine learning on difference image analysis: A comparison of methods for transient detection}",
      journal = {Astronomy and Computing},
         year = 2019,
        month = jul,
       volume = {28},
          eid = {100284},
        pages = {100284},
          doi = {10.1016/j.ascom.2019.05.002},
archivePrefix = {arXiv},
       eprint = {1812.10518},
 primaryClass = {astro-ph.IM},
       adsurl = {https://ui.adsabs.harvard.edu/abs/2019A&C....2800284S}
}

@ARTICLE{Francis2022,
       author = {{Francis}, Logan and {Johnstone}, Doug and {Lee}, Jeong-Eun and {Herczeg}, Gregory J. and {Long}, Feng and {Mairs}, Steve and {Contreras Pe{\~n}a}, Carlos and {Moriarty-Schieven}, Gerald and {JCMT Transient Team}},
        title = "{Accretion Burst Echoes as Probes of Protostellar Environments and Episodic Mass Assembly}",
      journal = {\apj},
         year = 2022,
        month = sep,
       volume = {937},
       number = {1},
          eid = {29},
        pages = {29},
          doi = {10.3847/1538-4357/ac8a9e},
archivePrefix = {arXiv},
       eprint = {2208.13568},
 primaryClass = {astro-ph.SR},
       adsurl = {https://ui.adsabs.harvard.edu/abs/2022ApJ...937...29F}
}
\end{multicols}

\end{document}